\documentclass{article}
\usepackage{spconf,amsmath,graphicx}
\usepackage{url}

\usepackage{amsmath,amsfonts}
\usepackage{algorithmic}
\usepackage{algorithm}
\usepackage{array}
\usepackage{textcomp}
\usepackage{stfloats}
\usepackage{url}
\usepackage{verbatim}
\usepackage{graphicx}
\usepackage{cite}

\usepackage{xcolor}
\usepackage{multirow}
\usepackage{makecell}
\usepackage{amsthm}
\usepackage{subfigure}
\usepackage{booktabs}

\usepackage{IEEEtrantools}

\title{Privacy-Aware Joint Source-Channel Coding for image transmission based on Disentangled Information Bottleneck}
\name{Lunan Sun\IEEEauthorrefmark{1}, Caili Guo\IEEEauthorrefmark{1}, Mingzhe Chen\IEEEauthorrefmark{2} and Yang Yang\IEEEauthorrefmark{1}.\thanks{This work was supported by the National Natural Science Foundation of China (62371070) and the Beijing Natural Science Foundation (L222043).}}
\address{\IEEEauthorrefmark{1}Beijing University of Posts and Telecommunications, China\\
\IEEEauthorrefmark{2}University of Miami, USA}
\begin{document}
%
\maketitle

\begin{abstract}
Current privacy-aware joint source-channel coding (JSCC) works aim at avoiding private information transmission by adversarially training the JSCC encoder and decoder under specific signal-to-noise ratios (SNRs) of eavesdroppers. However, these approaches incur additional computational and storage requirements as multiple neural networks must be trained for various eavesdroppers' SNRs to determine the transmitted information. To overcome this challenge, we propose a novel privacy-aware JSCC for image transmission based on disentangled information bottleneck (DIB-PAJSCC). In particular, we derive a novel disentangled information bottleneck objective to disentangle private and public information. Given the separate information, the transmitter can transmit only public information to the receiver while minimizing reconstruction distortion. Since DIB-PAJSCC transmits only public information regardless of the eavesdroppers' SNRs, it can eliminate additional training adapted to eavesdroppers' SNRs. Experimental results show that DIB-PAJSCC can reduce the eavesdropping accuracy on private information by up to 20\% compared to existing methods.
\end{abstract}
\section{Introduction}\label{sec:Introduction}
Joint source and channel coding (JSCC) has attracted increasing attention as a means to achieve reliable data transmission. Driven by the rapid advancements in artificial intelligence\cite{chen2021distributed, Yang2022Semantic}, deep learning (DL) based JSCC approaches have been proposed. In particular, due to the larger dimensions of images compared to speech and text data, there exists more potential redundancy in images. Therefore, it is challenging but worthy to design DL-based JSCC for image transmission. The DL-based JSCC for image transmission possesses appealing properties compared with SSCC, such as higher image restoration quality\cite{bourtsoulatze2019deep,  kurka2020deepjscc, sun2023adaptive} and robustness against bit errors \cite{choi2019neural,song2020infomax}. The aforementioned studies only consider the image reconstruction quality while ignoring the potential privacy leakage during image transmission \cite{bourtsoulatze2019deep, choi2019neural, sun2023adaptive,kurka2020deepjscc,song2020infomax}. To enhance security, recent studies on JSCC for image transmission have taken into account the privacy awareness\cite{Marchioro2020Adversarial, Erdemir2022Privacy} by adversarially training JSCC encoder and decoder to avoid the private information transmission.

These privacy-aware JSCC methods cannot separate private and public information in the transmitted codewords due to the inexplicability of neural networks. Therefore, they need to train multiple neural networks for different channel conditions, i.e., signal noise ratios (SNRs) to determine the information contained in the transmitted codewords, leading to extra computational resources and storage. In addition, they assume that eavesdroppers' SNRs used during neural network optimization are the same as those used during inference, which may lead to serious performance degradation due to SNR mismatch between optimization and inference\cite{xu2021wireless}.

To solve these challenges, we can separate private information and public information in the transmitted codewords, thus avoiding the private information transmission by transmitting only public information. Recently, the authors in \cite{tishby2000information} introduced an information-theoretic principle for neural networks, which enables neural networks to divide the image data that will be transmitted to the receiver into the target-relevant and target-irrelevant information\cite{pan2021disentangled}. However, the information-theoretic principle in \cite{tishby2000information} is designed for image classification and does not consider communication over channel. Thus, it is not suitable for the considered scenario where images are transmitted over channel and recovered by the receiver. Therefore, we propose a novel privacy-aware JSCC for image transmission based on disentangled information bottleneck (DIB-PAJSCC), which disentangles private and public information. The proposed approach can avoid private information transmission for eavesdroppers with different SNRs without additional training. In particular, we design a new disentangled IB objective that aims at minimizing the reconstruction distortion without transmitting private information. A new tractable estimation of the proposed objective is derived and used as the loss function of JSCC. We compare our approach with state-of-the-art privacy-aware JSCC methods via extensive experiments. Experimental results show that the proposed approach can significantly reduce eavesdropping accuracy on private information by up to 20\% compared to existing privacy-aware JSCC.

\vspace{-2mm}
\section{System Model}
\begin{figure}[t]
\centering{
\includegraphics[width=1\columnwidth]{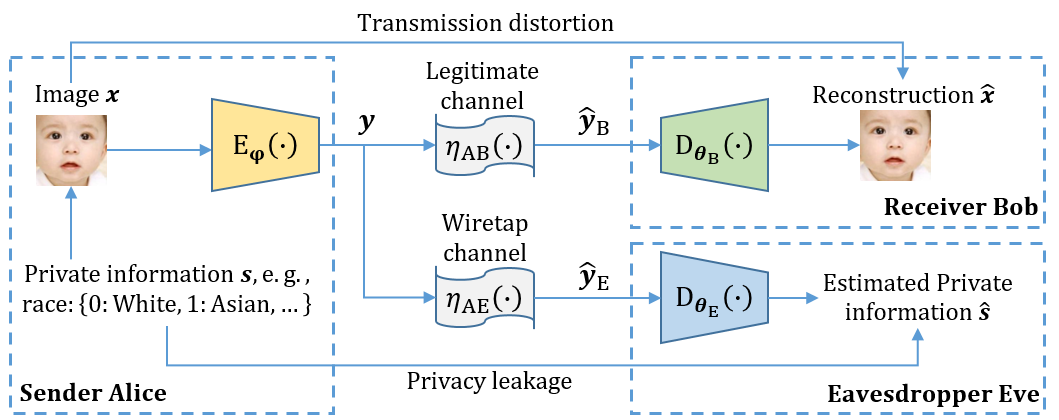}
}
\caption{\label{fig:illustration}An illustration of privacy-aware JSCC for image transmission, where the eavesdropper Eve tries to estimate private information $\boldsymbol{s}$ according to the transmitted codeword $\boldsymbol{y}$.}
\vspace{-3mm}
\end{figure}

\vspace{-2mm}
\label{sec:Problem}
As shown in Fig. \ref{fig:illustration}, we consider a system where a sender Alice transmits an image $\boldsymbol{x} \in {\mathbb{R}^{N}}$ with size $N$ to a legitimate receiver Bob, where $\mathbb{R}$ represents the set of real numbers. $\boldsymbol{x}$ contains some private information ${\boldsymbol{s}}$. Alice encodes $\boldsymbol{x}$ into a codeword $\boldsymbol{y} \in {\mathbb{R}^{M}}$, where $M$ represents the length of $\boldsymbol{y}$. The encoding function ${{\rm{E}}_{\boldsymbol{\varphi }}}:{{\mathbb{R}^{N}}} \to {{\mathbb{R}^{M}}}$ is an encoder neural network parameterized by $\boldsymbol{\varphi } $. $\boldsymbol{y}$ is transmitted over a noisy channel ${\eta _{{\rm{AB}}}}:{\mathbb{R}^{M}} \to {\mathbb{R}^{M}}$. The noisy codeword received by Bob is ${{{\boldsymbol{\hat y}}}_{\rm{B}}} = {\boldsymbol{y}} + {{\boldsymbol{z}}_{\rm{B}}}$, where ${{\boldsymbol{z}}_{\rm{B}}} \sim \mathcal{N}\left( {0,\sigma _{\rm{B}}^2{\boldsymbol{I}}} \right)$ represents the additive white Gaussian noise at Bob. Meanwhile, an external eavesdropper Eve can access $\boldsymbol{y}$ via an eavesdropping channel ${\eta _{{\rm{AE}}}}:{\mathbb{R}^{M}} \to {\mathbb{R}^{M}}$. The noisy codeword received by Eve is ${{{\boldsymbol{\hat y}}}_{\rm{E}}} = {\boldsymbol{y}} + {{\boldsymbol{z}}_{\rm{E}}}$, where ${{\boldsymbol{z}}_{\rm{E}}} \sim \mathcal{N}\left( {0,\sigma _{\rm{E}}^2{\boldsymbol{I}}} \right)$ represents the additive white Gaussian noise at Eve. Bob decodes the noisy codeword ${\boldsymbol{\hat y }}_{\rm{B}}$ into reconstructed image $\boldsymbol{\hat x } \in {\mathbb{R}^{N}}$. The decoding function ${{\rm{D}}_{\boldsymbol{\theta}_{\rm{B}}} }:{\mathbb{R}^{M}} \to {\mathbb{R}^{N}}$ is a decoder neural network with parameters $\boldsymbol{\theta}_{\rm{B}}$. The reconstructed image is ${\boldsymbol{\hat x}} = {{\rm{D}}_{{{\boldsymbol{\theta }}_{\rm{B}}}}}\left( {{{{\boldsymbol{\hat y}}}_{\rm{B}}}} \right)$. Meanwhile, Eve estimates private information $\boldsymbol {s}$ in $\boldsymbol {x}$ from the received codeword ${\boldsymbol{\hat y }}_{\rm{E}}$ using its own neural network with parameter ${\boldsymbol{\theta_{\rm{E}}}}$, ${{\rm{D}}_{\boldsymbol{\theta_{\rm{E}}}} }:{\mathbb{R}^{M}} \to {\left\{ {0,1} \right\}^S}$. The estimated private information at Eve is ${\boldsymbol{\hat s}} = {{\rm{D}}_{{{\boldsymbol{\theta }}_{\rm{E}}}}}\left( {{{{\boldsymbol{\hat y}}}_{\rm{E}}}} \right)$.

The goal of the considered system is to determine the encoder parameters ${\boldsymbol{\varphi }}$ and decoder parameters ${{\boldsymbol{\theta }}_{\rm{B}}}$  that minimize the average reconstruction error between $\boldsymbol{x}$ and $\boldsymbol{\hat x }$ while guaranteeing the estimated private information ${{\boldsymbol{\hat s}}}$ at Eve to be different from the private information $\boldsymbol{s}$ in $\boldsymbol{x}$.
\vspace{-3mm}
\section{DIB-PAJSCC}
\vspace{-3mm}
\label{sec:method}
In the considered privacy-aware JSCC, we apply the disentangled IB objective to disentangle $\boldsymbol{y }$ into the public subcodeword ${{\boldsymbol{y}}_{\rm{t}}} \in {\mathbb{R}^{{M_{\rm{t}}}}}$ and the private subcodeword ${{\boldsymbol{y}}_{\rm{s}}} \in {\mathbb{R}^{{M_{\rm{s}}}}}$, which are independent with each other. Therefore, we divide ${{\rm{E}}_{\boldsymbol{\varphi}} }\left(  \cdot  \right)$ into two components, the public encoder with parameters ${{{\boldsymbol{\varphi }}_{\rm{t}}}}$, $f_{{{\boldsymbol{\varphi }}_{\rm{t}}}}:{{\mathbb{R}}^{N}} \to {{\mathbb{R}}^{M_{\rm{t}}}}$ and the private encoder with parameters ${{{\boldsymbol{\varphi }}_{\rm{s}}}}$, ${f_{{{\boldsymbol{\varphi }}_{\rm{s}}}}}:{{\mathbb{R}}^{N}} \to {{\mathbb{R}}^{M_{\rm{s}}}}$, where $M = {M_{\rm{t}}} + {M_{\rm{s}}}$ and ${\boldsymbol{y}} = {\rm{concat}}\left[ {{{\boldsymbol{y}}_{\rm{t}}},{{\boldsymbol{y}}_{\rm{s}}}} \right]$. The disentangled IB objective for separating ${{\boldsymbol{y}}_{\rm{t}}}$ and ${{\boldsymbol{y}}_{\rm{s}}}$ as well as minimizing the reconstruction error is
\begin{equation}
\label{DIB-JSCC-1}
\mathop {{\rm{min}}}\limits_{{{\boldsymbol{\varphi }}_{\rm{s}}},{{\boldsymbol{\varphi }}_{\rm{t}}},{{\boldsymbol{\theta }}_{\rm{B}}}} {{\rm{E}}_{p\left( {{\boldsymbol{x,s}}} \right)}}\left( {d\left( {{\boldsymbol{x}};{\boldsymbol{\hat x}}} \right)} \right) + \alpha I\left( {{{\boldsymbol{y}}_{\rm{t}}};{{\boldsymbol{y}}_{\rm{s}}}} \right) - \beta I\left( {{{\boldsymbol{y}}_{\rm{s}}};{\boldsymbol{s}}} \right),
\end{equation}
where $\alpha $ and $\beta$ are hyperparameters, $d\left(  \cdot  \right)$ is the mean squared error (MSE) distortion of image transmission, $I\left( {{{\boldsymbol{y}}_{\rm{t}}};{{\boldsymbol{y}}_{\rm{s}}}} \right)$ is the mutual information between ${{{\boldsymbol{y}}_{\rm{t}}}}$ and ${{{\boldsymbol{y}}_{\rm{s}}}}$, and $I\left( {{{\boldsymbol{y}}_{\rm{s}}};{\boldsymbol{s}}} \right)$ is the mutual information between ${{{\boldsymbol{y}}_{\rm{s}}}}$ and ${\boldsymbol{s}}$. The first term in (\ref{DIB-JSCC-1}) is used to minimize the image transmission error when separating ${{{\boldsymbol{y}}_{\rm{t}}}}$ and ${{{\boldsymbol{y}}_{\rm{s}}}}$. The second term is used to compress the information between ${{\boldsymbol{y}}_{\rm{s}}}$ and ${{\boldsymbol{y}}_{\rm{t}}}$ thus encouraging the independence between ${{\boldsymbol{y}}_{\rm{s}}}$ and ${{\boldsymbol{y}}_{\rm{t}}}$. The third term is used to preserve the information related to $\boldsymbol{s}$ in ${{\boldsymbol{y}}_{\rm{s}}}$. By jointly minimizing $I\left( {{{\boldsymbol{y}}_{\rm{t}}};{{\boldsymbol{y}}_{\rm{s}}}} \right)$ and maximizing $I\left( {{{\boldsymbol{y}}_{\rm{s}}};{\boldsymbol{s}}} \right)$, the private information in ${{{\boldsymbol{y}}_{\rm{t}}}}$ is removed, and the public information is stored only in ${{{\boldsymbol{y}}_{\rm{t}}}}$. This ensures that ${{\boldsymbol{y}}_{\rm{t}}}$ contains no private information, and can be directly transmitted over the channel. Since (\ref{DIB-JSCC-1}) simultaneously compresses the private information in ${{{\boldsymbol{y}}_{\rm{t}}}}$ and reducing the reconstruction distortion, we refer (\ref{DIB-JSCC-1}) as disentangled IB objective. However, (\ref{DIB-JSCC-1}) still cannot be applied to privacy-aware JSCC, since $I\left( {{{\boldsymbol{y}}_{\rm{t}}};{{\boldsymbol{y}}_{\rm{s}}}} \right)$ and $I\left( {{{\boldsymbol{y}}_{\rm{s}}};{\boldsymbol{s}}} \right)$ in (\ref{DIB-JSCC-1}) are intractable due to the unknown $p\left( {{{\boldsymbol{y}}_{\rm{t}}},{{\boldsymbol{y}}_{\rm{s}}}} \right)$, $p\left( {{{\boldsymbol{y}}_{\rm{s}}}} \right)$ and $p\left( {{{\boldsymbol{y}}_{\rm{t}}}} \right)$. Therefore, we next derive a variational lower bound on $I\left( {{{\boldsymbol{y}}_{\rm{s}}};{\boldsymbol{s}}} \right)$ and an estimation of $I\left( {{{\boldsymbol{y}}_{\rm{t}}};{{\boldsymbol{y}}_{\rm{s}}}} \right)$ instead.

In particular, instead of maximizing the true value of $I\left( {{{\boldsymbol{y}}_{\rm{s}}};{\boldsymbol{s}}} \right)$, we maximize its lower bound. According to the definition of mutual information and entropy, we have\cite{alemi2016deep}
\begin{equation}
\label{DIB-JSCC-B-1}
\begin{aligned}
I\left( {{{\boldsymbol{y}}_{\rm{s}}};{\boldsymbol{s}}} \right) & = H\left( {\boldsymbol{s}} \right) - H\left( {{\boldsymbol{s}}|{{\boldsymbol{y}}_{\rm{s}}}} \right) \ge  - H\left( {{\boldsymbol{s}}|{{\boldsymbol{y}}_{\rm{s}}}} \right)\\
& = {\mathbb{E}_{p\left( {{{\boldsymbol{y}}_{\rm{s}}},{\boldsymbol{s}}} \right)}}\left( {\log q\left( {{\boldsymbol{s}}|{{\boldsymbol{y}}_{\rm{s}}}} \right)} \right) + \underbrace {{\mathbb{E}_{p\left( {{{\boldsymbol{y}}_{\rm{s}}},{\boldsymbol{s}}} \right)}}\left( {\log \frac{{p\left( {{\boldsymbol{s}}|{{\boldsymbol{y}}_{\rm{s}}}} \right)}}{{q\left( {{\boldsymbol{s}}|{{\boldsymbol{y}}_{\rm{s}}}} \right)}}} \right)}_{{D_{{\rm{KL}}}}\left[ {p\left( {{\boldsymbol{s}}|{{\boldsymbol{y}}_{\rm{s}}}} \right)||q\left( {{\boldsymbol{s}}|{{\boldsymbol{y}}_{\rm{s}}}} \right)|} \right]  \ge 0} \\
& \ge {\mathbb{E}_{p\left( {{{\boldsymbol{y}}_{\rm{s}}},{\boldsymbol{s}}} \right)}}\left( {\log q\left( {{\boldsymbol{s}}|{{\boldsymbol{y}}_{\rm{s}}}} \right)} \right),
\end{aligned} 
\end{equation}
where $H({\boldsymbol{s}})$ is the entropy of $\boldsymbol{s}$, $H({\boldsymbol{s}}|{{{{\boldsymbol{y}}_{\rm{s}}}}})$ is the conditional entropy of $\boldsymbol{s}$ given ${{{\boldsymbol{y}}_{\rm{s}}}}$, ${q\left( {\boldsymbol{s}|{{\boldsymbol{y}}_{\rm{s}}}} \right)}$ is the variational approximation of the true posterior ${p\left( {\boldsymbol{s}|{{\boldsymbol{y}}_{\rm{s}}}} \right)}$. ${{E_{p\left( {{{\boldsymbol{y}}_{\rm{s}}},{\boldsymbol{s}}} \right)}}\left( {\log \frac{{p\left( {{\boldsymbol{s}}|{{\boldsymbol{y}}_{\rm{s}}}} \right)}}{{q\left( {{\boldsymbol{s}}|{{\boldsymbol{y}}_{\rm{s}}}} \right)}}} \right)}$ in the second row of (\ref{DIB-JSCC-B-1}) is the KL divergence between ${q\left( {\boldsymbol{s}|{{\boldsymbol{y}}_{\rm{s}}}} \right)}$ and ${p\left( {\boldsymbol{s}|{{\boldsymbol{y}}_{\rm{s}}}} \right)}$ and is larger than $0$. A classifier ${{\rm{C}}_{\boldsymbol{\gamma }}}:{\mathbb{R}^{M}} \to {\left\{ {0,1} \right\}^S}$ with parameters $\boldsymbol{\gamma }$ is applied to denote ${q\left( {\boldsymbol{s}|{{\boldsymbol{y}}_{\rm{s}}}} \right)}$. To make ${q\left( {{\boldsymbol{s}}|{{\boldsymbol{y}}_{\rm{s}}}} \right)}$ close to ${p\left( {{\boldsymbol{s}}|{{\boldsymbol{y}}_{\rm{s}}}} \right)}$, we optimize ${\boldsymbol{\gamma }}$ to minimize the cross entropy between ${{\rm{C}}_{\boldsymbol{\gamma }}}\left( {{{\boldsymbol{y}}_{\rm{s}}}} \right)$ and $p\left( {{\boldsymbol{s}}|{{\boldsymbol{y}}_{\rm{s}}}} \right)$. The trained ${{\rm{C}}_{\boldsymbol{\gamma }}}\left( {{{\boldsymbol{y}}_{\rm{s}}}} \right)$ is denoted as ${q\left( {{\boldsymbol{s}}|{{\boldsymbol{y}}_{\rm{s}}}} \right)}$. Since there exists a Markov chain ${\boldsymbol{s}} \leftrightarrow {\boldsymbol{x}} \leftrightarrow {{\boldsymbol{y}}_{\rm{s}}}$, $p\left( {{{\boldsymbol{y}}_{\rm{s}}},{\boldsymbol{s}}} \right) = p\left( {{\boldsymbol{x,s}}} \right)p\left( {{{\boldsymbol{y}}_{\rm{s}}}|{\boldsymbol{x}}} \right)$. The variational lower bound on $I\left( {{{\boldsymbol{y}}_{\rm{s}}};{\boldsymbol{s}}} \right)$ is estimated as
\begin{equation}
\label{DIB-JSCC-B-2}
I\left( {{{\boldsymbol{y}}_{\rm{s}}};{\boldsymbol{s}}} \right) \ge {\mathbb{E}_{p\left( {{\boldsymbol{x,s}}} \right)}}{\mathbb{E}_{p\left( {{{\boldsymbol{y}}_{\rm{s}}}|{\boldsymbol{x}}} \right)}}\left[ {\log {{\rm{C}}_{\boldsymbol{\gamma }}}\left( {{{\boldsymbol{y}}_{\rm{s}}}} \right)} \right]. 
\end{equation}
As we use a deterministic ${f_{{{\boldsymbol{\varphi }}_{\rm{s}}}}}$, $p\left( {{{\boldsymbol{y}}_{\rm{s}}}|{\boldsymbol{x}}} \right)$ can be regarded as a Dirac-delta function, i.e., 
\begin{equation}
\label{DIB-JSCC-B-3}
p\left( {{{\boldsymbol{y}}_{\rm{s}}}|{\boldsymbol{x}}} \right) =
\begin{cases}
\begin{aligned}
1 & \quad \text{if } {{\boldsymbol{y}}_{\rm{s}}} = {f_{{{\boldsymbol{\varphi }}_{\rm{s}}}}}\left( {\boldsymbol{x}} \right) \\
0 & \quad \text{else}
\end{aligned}
\end{cases}.
\end{equation}
Replacing $p\left( {{{\boldsymbol{y}}_{\rm{s}}}|{\boldsymbol{x}}} \right)$ in (\ref{DIB-JSCC-B-2})
with (\ref{DIB-JSCC-B-3}), the variational lower bound on $I\left( {{{\boldsymbol{y}}_{\rm{s}}};{\boldsymbol{s}}} \right)$ can be calculated as
\begin{equation}
\label{DIB-JSCC-B-4}
I\left( {{{\boldsymbol{y}}_{\rm{s}}};{\boldsymbol{s}}} \right) \ge {\mathbb{E}_{p\left( {{\boldsymbol{x}},{\boldsymbol{s}}} \right)}}\left( {\log {{\rm{C}}_{\boldsymbol{\gamma }}}\left( {{f_{{{\boldsymbol{\varphi }}_{\rm{s}}}}}\left( {\boldsymbol{x}} \right)} \right)} \right).
\end{equation}

By maximizing the variational lower bound on $I\left( {{{\boldsymbol{y}}_{\rm{s}}};{\boldsymbol{s}}} \right)$, the private information can be converged in ${{\boldsymbol{y}}_{\rm{s}}}$. It is also crucial to minimize $I\left( {{{\boldsymbol{y}}_{\rm{t}}};{{\boldsymbol{y}}_{\rm{s}}}} \right)$ to enforce independence between ${{\boldsymbol{y}}_{\rm{s}}}$ and ${{\boldsymbol{y}}_{\rm{t}}}$ and prevent any private information from leaking into ${{\boldsymbol{y}}_{\rm{t}}}$. However, minimizing $I\left( {{{\boldsymbol{y}}_{\rm{t}}};{{\boldsymbol{y}}_{\rm{s}}}} \right)$ is intractable since both ${p\left( {{{\boldsymbol{y}}_{\rm{t}}},{{\boldsymbol{y}}_{\rm{s}}}} \right)}$ and ${p\left( {{{\boldsymbol{y}}_{\rm{t}}}} \right)p\left( {{{\boldsymbol{y}}_{\boldsymbol{s}}}} \right)}$ involve mixtures with a large number of components and are intractable. Therefore, we estimate $I\left( {{{\boldsymbol{y}}_{\rm{t}}};{{\boldsymbol{y}}_{\rm{s}}}} \right)$ and minimize its estimation instead. We first sample several ${\boldsymbol{y}} = {\rm{concat}}\left[ {{{\boldsymbol{y}}_{\rm{t}}},{{\boldsymbol{y}}_{\rm{s}}}} \right]$. Denote $\tau \left( {{{\boldsymbol{y}}_{\rm{t}}},{{\boldsymbol{y}}_{\rm{s}}}} \right)$ as the probability that ${{{\boldsymbol{y}}_{\rm{t}}}}$ is interdependent with ${{{\boldsymbol{y}}_{\rm{s}}}}$. If ${{{\boldsymbol{y}}_{\rm{t}}}}$ and ${{{\boldsymbol{y}}_{\rm{s}}}}$ are sampled from ${p\left( {{{\boldsymbol{y}}_{\rm{t}}}} \right)p\left( {{{\boldsymbol{y}}_{\boldsymbol{s}}}} \right)}$,  we have $\tau \left( {{{\boldsymbol{y}}_{\rm{t}}},{{\boldsymbol{y}}_{\rm{s}}}} \right) = 0$. If ${{{\boldsymbol{y}}_{\rm{t}}}}$ and ${{{\boldsymbol{y}}_{\rm{s}}}}$ are sampled from ${p\left( {{{\boldsymbol{y}}_{\rm{t}}},{{\boldsymbol{y}}_{\rm{s}}}} \right)}$, we have $\tau \left( {{{\boldsymbol{y}}_{\rm{t}}}, {{\boldsymbol{y}}_{\rm{s}}}} \right) = 1$. Then, $I\left( {{{\boldsymbol{y}}_{\rm{t}}};{{\boldsymbol{y}}_{\rm{s}}}} \right)$ can be expressed as\cite{pmlr-v80-kim18b, NEURIPS2018_1ee3dfcd, Chen_2021_ICCV}
\begin{equation}
\label{DIB-JSCC-C-2}
\begin{aligned}
I\left( {{{\boldsymbol{y}}_{\rm{t}}};{{\boldsymbol{y}}_{\rm{s}}}} \right) & = {\mathbb{E}_{p\left( {{{\boldsymbol{y}}_{\rm{t}}},{{\boldsymbol{y}}_{\rm{s}}}} \right)}}\left( {\log \frac{{p\left( {{{\boldsymbol{y}}_{\rm{t}}},{{\boldsymbol{y}}_{\rm{s}}}} \right)}}{{p\left( {{{\boldsymbol{y}}_{\rm{t}}}} \right)p\left( {{{\boldsymbol{y}}_{\boldsymbol{s}}}} \right)}}} \right)
\\& = {\mathbb{E}_{p\left( {{{\bf{y}}_{\rm{t}}},{{\bf{y}}_{\rm{s}}}} \right)}}\left( {\log \frac{{p\left( {\tau \left( {{{\bf{y}}_{\rm{t}}},{{\bf{y}}_{\rm{s}}}} \right) = 1} \right)}}{{1 - p\left( {\tau \left( {{{\bf{y}}_{\rm{t}}},{{\bf{y}}_{\rm{s}}}} \right) = 1} \right)}}} \right).
\end{aligned}
\end{equation}
From (\ref{DIB-JSCC-C-2}), the estimation of $I\left( {{{\boldsymbol{y}}_{\rm{t}}};{{\boldsymbol{y}}_{\rm{s}}}} \right)$ requires only the probability ${p\left( {\tau \left( {{{\boldsymbol{y}}_{\rm{t}}},{{\boldsymbol{y}}_{\rm{s}}}} \right) = 1} \right)}$. However, directly estimating ${p\left( {\tau \left( {{{\boldsymbol{y}}_{\rm{t}}},{{\boldsymbol{y}}_{\rm{s}}}} \right) = 1} \right)}$ using Monte Carlo does
not work due to high dimensions of ${{{\boldsymbol{y}}_{\rm{t}}}}$ and ${{{\boldsymbol{y}}_{\rm{s}}}}$\cite{pmlr-v80-kim18b}. Hence, we employ the density-ratio trick \cite{nguyen2010estimating} that involves a discriminator to approximate ${p\left( {\tau \left( {{{\boldsymbol{y}}_{\rm{t}}},{{\boldsymbol{y}}_{\rm{s}}}} \right) = 1} \right)}$. Denote the discriminator that consists of neural network with parameters ${\boldsymbol{\varepsilon }}$ as ${\rm{Di}}{{\rm{s}}_{\boldsymbol{\varepsilon }}}:{\mathbb{R}^{M}} \to {[0,1]^2}$.  The output of ${{\rm{Di}}{{\rm{s}}_{\boldsymbol{\varepsilon }}}\left( {{{\boldsymbol{y}}_{\rm{t}}},{{\boldsymbol{y}}_{\rm{s}}}} \right)}$ is treated as ${p\left( {\tau \left( {{{\boldsymbol{y}}_{\rm{t}}},{{\boldsymbol{y}}_{\rm{s}}}} \right) = 1} \right)}$. The samples from ${p\left( {{{\boldsymbol{y}}_{\rm{t}}},{{\boldsymbol{y}}_{\rm{s}}}} \right)}$ are obtained by first choosing ${\boldsymbol{x}}$ uniformly at random and then sampling from $p\left( {{{\boldsymbol{y}}_{\rm{t}}}|{\boldsymbol{x}}} \right)$ and $p\left( {{{\boldsymbol{y}}_{\rm{s}}}|{\boldsymbol{x}}} \right)$. The samples from ${p\left( {{{\boldsymbol{y}}_{\rm{t}}}} \right)p\left( {{{\boldsymbol{y}}_{\boldsymbol{s}}}} \right)}$ are obtained by first sampling from ${p\left( {{{\boldsymbol{y}}_{\rm{t}}},{{\boldsymbol{y}}_{\rm{s}}}} \right)}$ and then permuting ${{{\boldsymbol{y}}_{\rm{t}}}}$ and ${{{\boldsymbol{y}}_{\rm{s}}}}$ along the batch axis. Using the samples from ${\tau \left( {{{\boldsymbol{y}}_{\rm{t}}},{{\boldsymbol{y}}_{\rm{s}}}} \right) = 1}$ and ${\tau \left( {{{\boldsymbol{y}}_{\rm{t}}},{{\boldsymbol{y}}_{\rm{s}}}} \right) = 0}$, we train ${\rm{Di}}{{\rm{s}}_{\boldsymbol{\varepsilon }}}$ to distinguish samples from ${p\left( {{{\boldsymbol{y}}_{\rm{t}}},{{\boldsymbol{y}}_{\rm{s}}}} \right)}$ and ${p\left( {{{\boldsymbol{y}}_{\rm{t}}}} \right)p\left( {{{\boldsymbol{y}}_{\boldsymbol{s}}}} \right)}$. The loss function of  ${\rm{Di}}{{\rm{s}}_{\boldsymbol{\varepsilon }}}$ is
\begin{equation}
\label{DIB-JSCC-C-4}
\mathop {\min }\limits_{\boldsymbol{\varepsilon }} \log {\rm{Di}}{{\rm{s}}_{\boldsymbol{\varepsilon }}}\left( {{{\boldsymbol{y}}_{\rm{t}}},{{\boldsymbol{y}}_{\rm{s}}}} \right) + \log \left( {1 - {\rm{Di}}{{\rm{s}}_{\boldsymbol{\varepsilon }}}\left( {{{{\boldsymbol{\tilde y}}}_{\rm{t}}},{{{\boldsymbol{\tilde y}}}_{\rm{s}}}} \right)} \right),
\end{equation}
where ${{{{\boldsymbol{\tilde y}}}_{\rm{t}}}}$ and ${{{{\boldsymbol{\tilde y}}}_{\rm{s}}}}$ are the results by randomly permuting ${{{\boldsymbol{y}}_{\rm{t}}}}$ and ${{{\boldsymbol{y}}_{\rm{s}}}}$ along the batch axis, respectively. By optimizing (\ref{DIB-JSCC-C-4}), the output of ${\rm{Di}}{{\rm{s}}_{\boldsymbol{\varepsilon }}}$ will be forced to 0 when ${{{\boldsymbol{y}}_{\rm{t}}}}$ and ${{{\boldsymbol{y}}_{\rm{s}}}}$ are 
independent and to 1 when ${{{\boldsymbol{y}}_{\rm{t}}}}$ and ${{{\boldsymbol{y}}_{\rm{s}}}}$ are dependent. After training, (\ref{DIB-JSCC-C-2}) can be expressed as
\begin{equation}
\label{DIB-JSCC-C-5}
I\left( {{{\boldsymbol{y}}_{\rm{t}}};{{\boldsymbol{y}}_{\rm{s}}}} \right) \approx {\mathbb{E}_{p\left( {{{\boldsymbol{y}}_{\rm{t}}},{{\boldsymbol{y}}_{\rm{s}}}} \right)}}\left( {\log \frac{{{\rm{Di}}{{\rm{s}}_{\boldsymbol{\varepsilon }}}\left( {{{\boldsymbol{y}}_{\rm{t}}},{{\boldsymbol{y}}_{\rm{s}}}} \right)}}{{1 - {\rm{Di}}{{\rm{s}}_{\boldsymbol{\varepsilon }}}\left( {{{\boldsymbol{y}}_{\rm{t}}},{{\boldsymbol{y}}_{\rm{s}}}} \right)}}} \right).
\end{equation}
Then, we can use (\ref{DIB-JSCC-C-5}) as the loss function to optimize ${{\boldsymbol{\varphi }}_{\rm{t}}}$. The proposed disentangled IB objective can be calculated by replacing $I\left( {{{\boldsymbol{y}}_{\rm{s}}};{\boldsymbol{s}}} \right)$ and $I\left( {{{\boldsymbol{y}}_{\rm{t}}};{{\boldsymbol{y}}_{\rm{s}}}} \right)$ in (\ref{DIB-JSCC-1}) with (\ref{DIB-JSCC-B-4}) and (\ref{DIB-JSCC-C-5}). However, we experimentally observe that when simultaneously training ${f_{{{\boldsymbol{\varphi }}_{\rm{s}}}}}$ and ${f_{{{\boldsymbol{\varphi }}_{\rm{t}}}}}$,  the encoder network will converge to a degenerated solution, where all information is encoded in ${{{\boldsymbol{y}}_{\rm{s}}}}$, whereas ${{{\boldsymbol{y}}_{\rm{t}}}}$ holds almost no information. To prevent this undesirable solution, we adopt a two-step training strategy \cite{Hadad2018two}. In the first step, ${f_{{{\boldsymbol{\varphi }}_{\rm{s}}}}}$ and ${{\rm{C}}_{\boldsymbol{\gamma }}}$ are jointly trained using (\ref{DIB-JSCC-B-4}) as the loss function to extract ${{{\boldsymbol{y}}_{\rm{s}}}}$ that contains private information. In the second step, ${f_{{{\boldsymbol{\varphi }}_{\rm{s}}}}}$ is freezed. ${f_{{{\boldsymbol{\varphi }}_{\rm{t}}}}}$ and ${{\rm{D}}_{{{\boldsymbol{\theta }}_{\rm{B}}}}}$ are jointly trained using (\ref{DIB-JSCC-C-5}) as loss function, followed by alternating training with ${\rm{Di}}{{\rm{s}}_{\boldsymbol{\varepsilon }}}$ using (\ref{DIB-JSCC-C-4}) as loss function in order to enable ${{{\boldsymbol{y}}_{\rm{t}}}}$ to capture public information. By training ${f_{{{\boldsymbol{\varphi }}_{\rm{s}}}}}$ in the first step, and freezing its parameters in the second step, ${f_{{{\boldsymbol{\varphi }}_{\rm{s}}}}}$ has a limited capacity since it ignores most of the public information and thus enabling ${f_{{{\boldsymbol{\varphi }}_{\rm{t}}}}}$ to extract public information. After training, ${{{\boldsymbol{y}}_{\boldsymbol{s}}}}$ are fixed to 0. ${{\rm{D}}_{{{\bf{\theta }}_{\rm{B}}}}}$ is further trained to reduce the reconstruction distortion for several epochs.

It is worth noting that the whole training process only requires the SNR of Bob, ${\rm{SN}}{{\rm{R}}_{{\rm{AB}}}}$, and does not require the SNR of Eve, ${\rm{SN}}{{\rm{R}}_{{\rm{AE}}}}$. Therefore, DIB-PAJSCC is effective under all ${\rm{SN}}{{\rm{R}}_{{\rm{AE}}}}$, which solves the issue that the current privacy-aware JSCC requires a specified ${\rm{SN}}{{\rm{R}}_{{\rm{AE}}}}$.

\vspace{-5mm}
\section{Experimental results}
\vspace{-3mm}
\label{sec:experiment}
In this section, we compare the performance of DIB-PAJSCC with adversarial privacy-aware JSCC\cite{Marchioro2020Adversarial, Erdemir2022Privacy}, referred to as Adv., in terms of eavesdropping accuracy on private information and reconstruction quality. The experiments are carried on the colored MNIST dataset\cite{lecun1998mnist} and the UTK Face dataset\cite{Zhang2017Age}. The color (total 10 categories) and the ethnicity (total 5 categories) are set as ${\boldsymbol{s}}$ for the colored MNIST and the UTK Face dataset.

\begin{figure*}[t!]
\centering 
\setlength{\abovecaptionskip}{0cm}
\setlength{\belowcaptionskip}{-0.5cm}
\subfigbottomskip=7pt 
\subfigcapskip=0pt 
\subfigure[The colored MNIST dataset.]{
\label{fig:snrAB1}
\includegraphics[width=.45\linewidth]{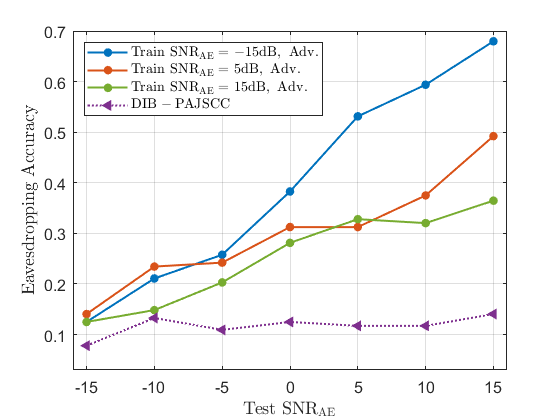}}
\subfigure[The UTK face dataset.]{
\label{fig:snrAB2}
\includegraphics[width=.45\linewidth]{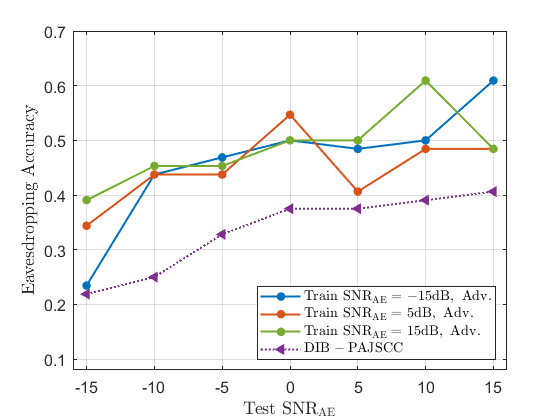}}
\vspace{-3mm}
\caption{The eavesdropping accuracy of Adv. and DIB-PAJSCC. DIB-PAJSCC requires to train only one neural network and outperforms Adv., which needs to train multiple neural networks for different assumptions of ${\rm{SN}}{{\rm{R}}_{{\rm{AE}}}}$.}
\vspace{-3mm}
\label{fig:snrAE}
\end{figure*} 

Figure \ref{fig:snrAE} shows the eavesdropping accuracy on private information of adversarial privacy-aware JSCC and DIB-PAJSCC under various test ${\rm{SN}}{{\rm{R}}_{{\rm{AE}}}}$. MSEs of all methods are kept close. From Fig. \ref{fig:snrAE}, we can observe that the eavesdropping accuracy of DIB-PAJSCC is always lower than that of the adversarial privacy-aware JSCC. When test ${\rm{SN}}{{\rm{R}}_{{\rm{AE}}}} = 15 {\rm{dB}}$, DIB-PAJSCC can reduce up to 20\% eavesdropping accuracy than the adversarial privacy-aware JSCC. This implies that DIB-PAJSCC has better robustness when there is an estimated error on ${\rm{SN}}{{\rm{R}}_{{\rm{AE}}}}$. On the colored MNIST dataset, the eavesdropping accuracy of DIB-PAJSCC is close to random guess (about 0.1 for 10 categories) and exhibits minimal variation when ${\rm{SN}}{{\rm{R}}_{{\rm{AE}}}}$ increases. However, the eavesdropping accuracy of adversarial privacy-aware JSCC increases obviously as ${\rm{SN}}{{\rm{R}}_{{\rm{AE}}}}$ increases. This is because the color information can be completely separated from other information. ${{\boldsymbol{y}}_{\rm{t}}}$ of DIB-PAJSCC contains no information about color, thus leading to eavesdropping accuracy close to that of a random guess. 

\begin{figure}[t!]
\centering{
\includegraphics[width=0.7\linewidth]{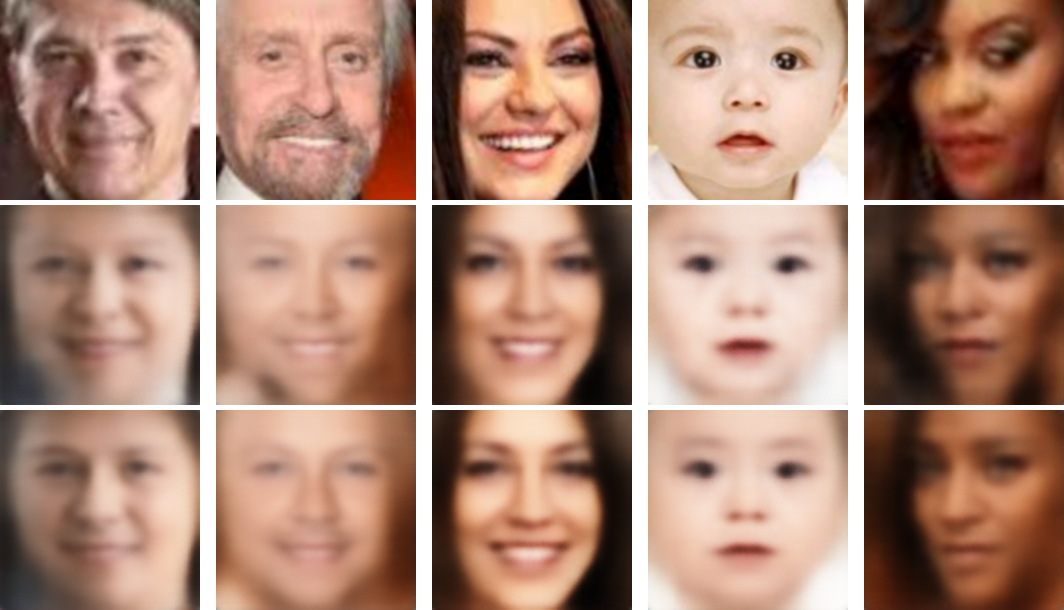}
}
\vspace{-2mm}
\caption{\label{fig:SSCC_Reconstructions}Images transmitted by Alice (top) and images recovered by Adv. (middle), recovered by DIB-PAJSCC (bottom).}
\vspace{-4mm}
\end{figure}
Figure \ref{fig:SSCC_Reconstructions} shows the visual reconstructions of adversarial privacy-aware JSCC and DIB-PAJSCC. From Fig. \ref{fig:SSCC_Reconstructions}, we can observe that the visual quality of adversarial privacy-aware JSCC and DIB-PAJSCC are similar since their reconstruction MSEs are close. Even though DIB-PAJSCC sacrifices a little reconstruction quality to defend against eavesdropping, the key facial information such as eye and nose positions is not damaged. This demonstrates that DIB-PAJSCC can preserve the semantic information irrelevant to privacy well.

\begin{figure}[t!]
\centering 
\setlength{\abovecaptionskip}{0cm}
\setlength{\belowcaptionskip}{-0.5cm}
\subfigbottomskip=7pt 
\subfigcapskip=0pt 
\subfigure[t-SNE of ${{\boldsymbol{y}}_{\rm{t}}}$.]{
\label{fig:TSNE-method1}
\includegraphics[width=.45\linewidth]{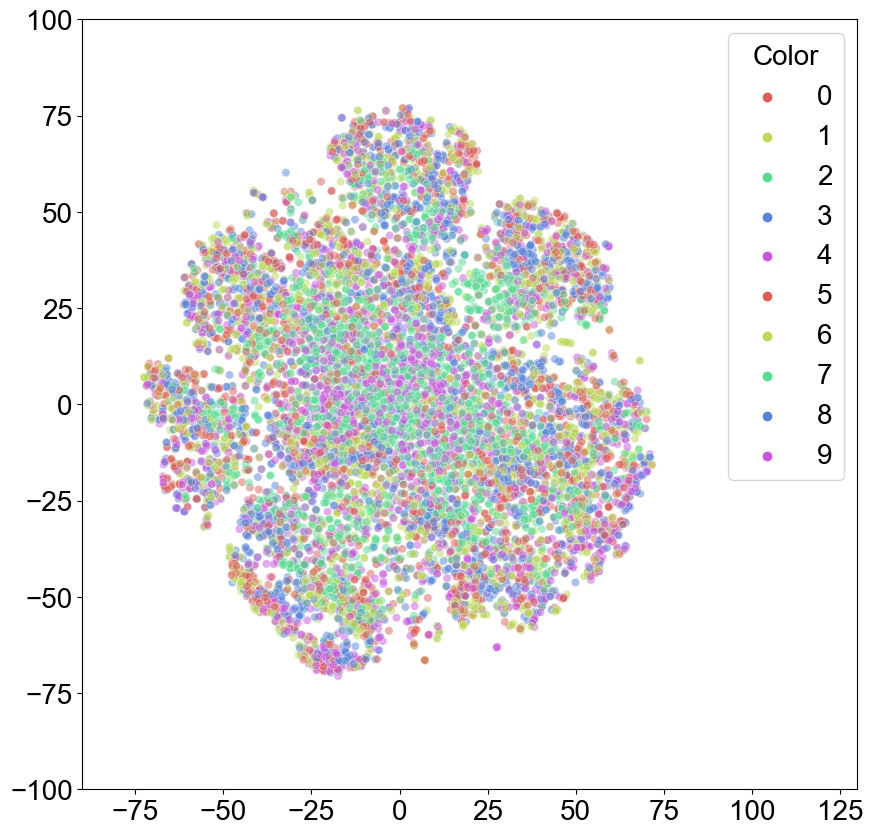}}
\subfigure[t-SNE of ${{\boldsymbol{y}}_{\rm{t}}}$.]{
\label{fig:TSNE-method2}
\includegraphics[width=.45\linewidth]{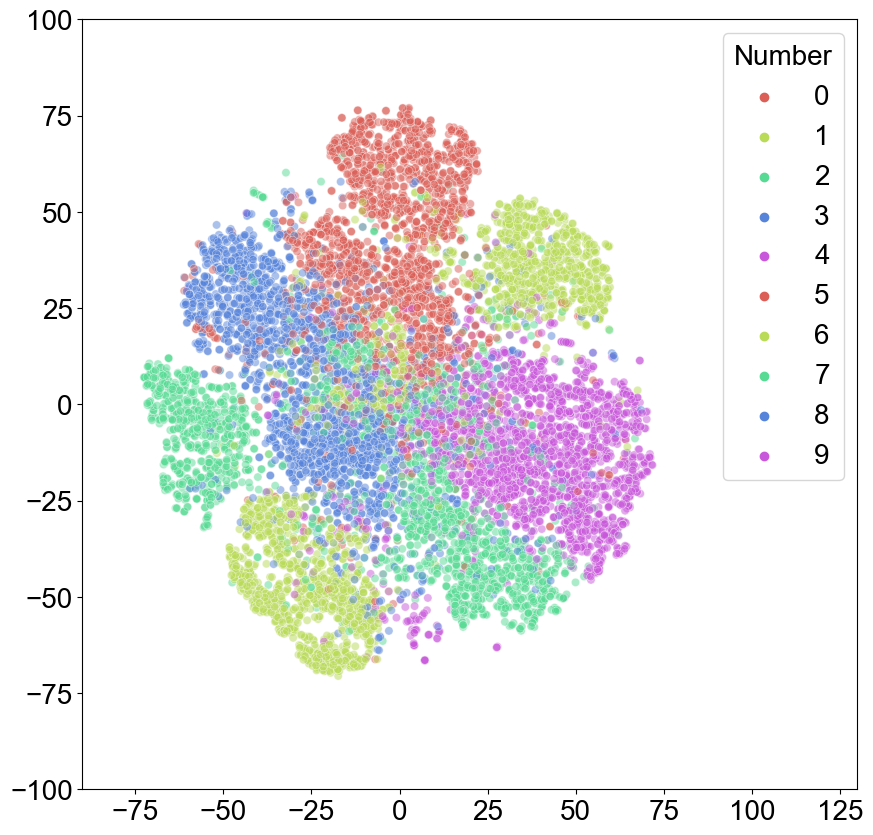}}
\subfigure[t-SNE of ${{\boldsymbol{y}}_{\rm{s}}}$.]{
\label{fig:TSNE-method3}
\includegraphics[width=.45\linewidth]{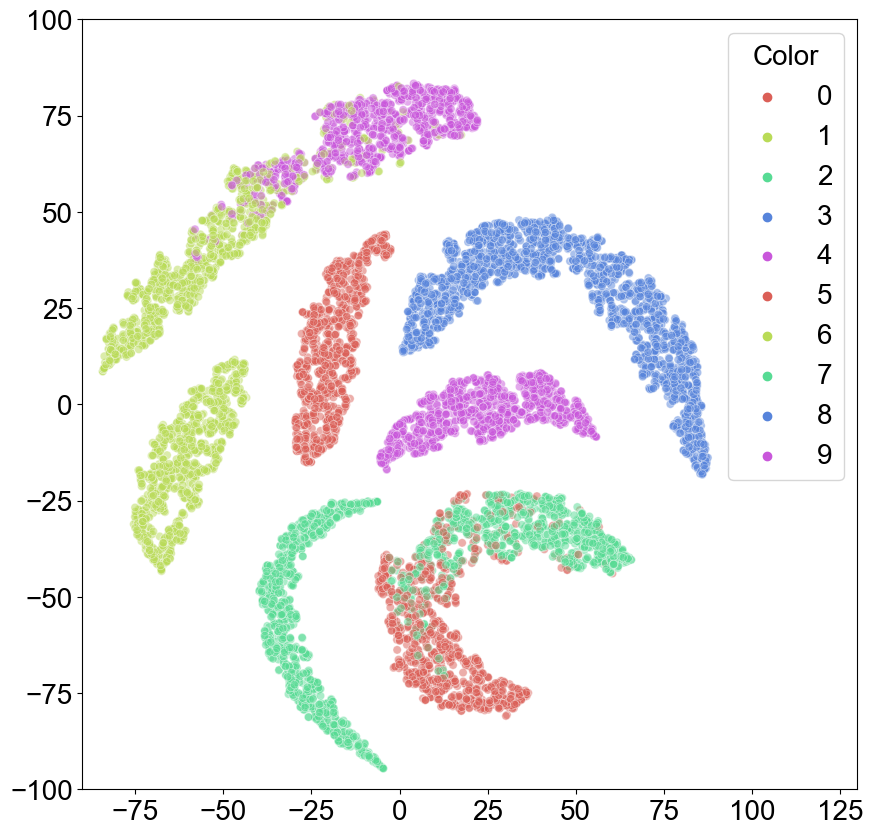}}
\subfigure[t-SNE of ${{\boldsymbol{y}}_{\rm{s}}}$.]{
\label{fig:TSNE-method4}
\includegraphics[width=.45\linewidth]{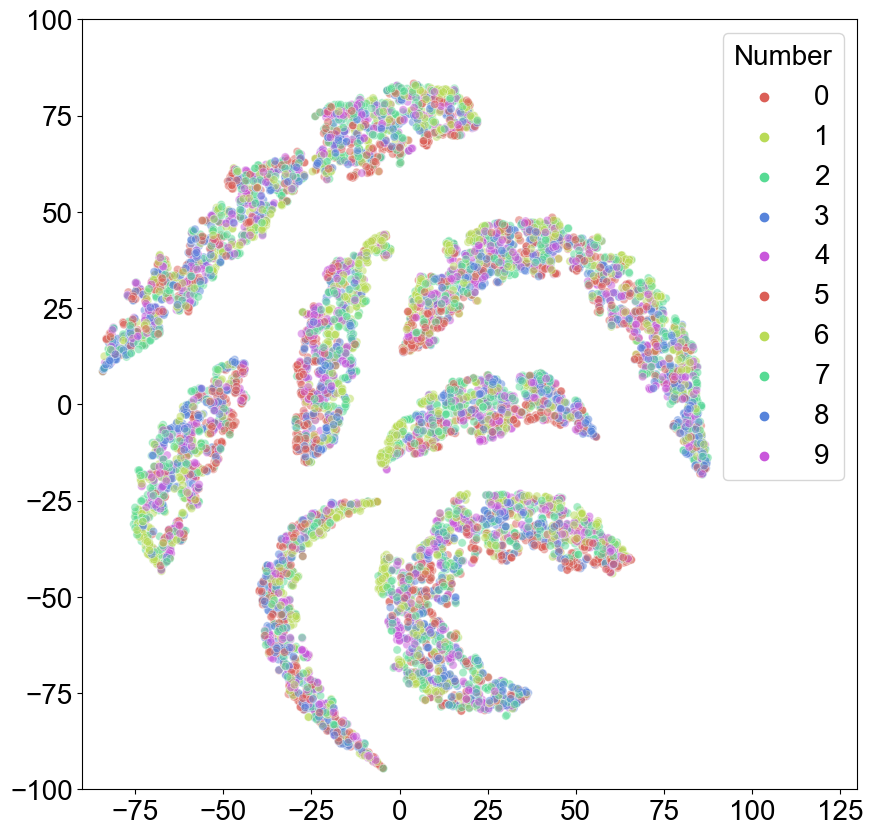}}
\vspace{-4mm}
\caption{t-SNE visualization of ${{\boldsymbol{y}}_{\rm{t}}}$ and ${{\boldsymbol{y}}_{\rm{s}}}$.}
\vspace{-5mm}
\label{fig:TSNE-method}
\end{figure}

Figure \ref{fig:TSNE-method} shows the 2-dimensional projections of ${{\boldsymbol{y}}_{\rm{t}}}$ and ${{\boldsymbol{y}}_{\rm{s}}}$ from the colored MNIST dataset. ${{\boldsymbol{y}}_{\rm{t}}}$ and ${{\boldsymbol{y}}_{\rm{s}}}$ are projected into a 2-dimensional space utilizing t-Distributed Stochastic Neighbor Embedding (t-SNE)\cite{van2008visualizing}. We also show the labels of each image with regard to the private information, i.e., color, and the public information, i.e., digits, to make it easier to investigate the clusters. From Fig. \ref{fig:TSNE-method1} and \ref{fig:TSNE-method2}, we can observe that the t-SNE projections of ${{\boldsymbol{y}}_{\rm{t}}}$ with different colors exhibit significant overlap, while the t-SNE projections of ${{\boldsymbol{y}}_{\rm{t}}}$ with different digits are separated well. This indicates that ${{\boldsymbol{y}}_{\rm{t}}}$ contains abundant digit-related information and is agnostic to the private information. From Fig. \ref{fig:TSNE-method3} and \ref{fig:TSNE-method4}, we can also observe that the t-SNE projections of ${{\boldsymbol{y}}_{\rm{s}}}$ with different colors are distinctly separated, while the t-SNE projections of ${{\boldsymbol{y}}_{\rm{s}}}$ with different digits are mixed together. This suggests that ${{\boldsymbol{y}}_{\rm{s}}}$ contains abundant color-related information while almost no digit-related information. This is because DIB-PAJSCC preserves as much private information in ${{\boldsymbol{y}}_{\rm{s}}}$ as possible, and removes the private information in ${{\boldsymbol{y}}_{\rm{t}}}$ at the same time. In addition, to guarantee the reconstruction quality, DIB-PAJSCC preserves the public information in ${{\boldsymbol{y}}_{\rm{t}}}$ instead of ${{\boldsymbol{y}}_{\rm{s}}}$, as ${{\boldsymbol{y}}_{\rm{t}}}$ is directly transmitted to Bob. Hence, DIB-PAJSCC is able to disentangle private and public information and preserve them in the proper subcodewords.

\vspace{-5mm}
\section{conclusion}
\vspace{-3mm}
\label{sec:conclusion}
In this work, we have proposed a DIB-PAJSCC scheme for image transmission, which can prevent privacy leakage caused by eavesdroppers with different SNRs without additional training. Specifically, we derived a tractable form of the disentangled IB objective for disentangling private and public information, and only public information is transmitted. Experimental results have shown that DIB-PAJSCC can significantly reduce privacy leakage and preserve public information well.

\bibliographystyle{IEEEbib}
\bibliography{refs}

\begin{thebibliography}{10}

\bibitem{chen2021distributed}
Mingzhe Chen, Deniz G{\"u}nd{\"u}z, Kaibin Huang, Walid Saad, Mehdi Bennis,
  Aneta~Vulgarakis Feljan, and H~Vincent Poor,
\newblock ``Distributed learning in wireless networks: Recent progress and
  future challenges,''
\newblock {\em {IEEE} J. Sel. Areas Commun.}, vol. 39, no. 12, pp. 3579--3605,
  Oct. 2021.

\bibitem{Yang2022Semantic}
Yang Yang, Caili Guo, Fangfang Liu, Chuanhong Liu, Lunan Sun, Qizheng Sun, and
  Jiujiu Chen,
\newblock ``Semantic communications with artificial intelligence tasks:
  Reducing bandwidth requirements and improving artificial intelligence task
  performance,''
\newblock {\em {IEEE} Ind. Electron. Mag.}, pp. 2--11, 2022, Early Access.

\bibitem{bourtsoulatze2019deep}
Eirina Bourtsoulatze, David~Burth Kurka, and Deniz G{\"u}nd{\"u}z,
\newblock ``Deep joint source-channel coding for wireless image transmission,''
\newblock {\em {IEEE} Trans. Cogn. Commun. Netw.}, vol. 5, no. 3, pp. 567--579,
  Sept. 2019.

\bibitem{kurka2020deepjscc}
David~Burth Kurka and Deniz G{\"u}nd{\"u}z,
\newblock ``Deep{JSCC}-f: Deep joint source-channel coding of images with
  feedback,''
\newblock {\em {IEEE} J. Sel. Areas Inf. Theory}, vol. 1, no. 1, pp. 178--193,
  May 2020.

\bibitem{sun2023adaptive}
Lunan Sun, Yang Yang, Mingzhe Chen, Caili Guo, Walid Saad, and H~Vincent Poor,
\newblock ``Adaptive information bottleneck guided joint source and channel
  coding for image transmission,''
\newblock {\em {IEEE} J. Sel. Areas Commun.}, vol. 41, no. 8, pp. 2628--2644,
  Aug. 2023.

\bibitem{choi2019neural}
Kristy Choi, Kedar Tatwawadi, Aditya Grover, Tsachy Weissman, and Stefano
  Ermon,
\newblock ``Neural joint source-channel coding,''
\newblock in {\em Proc. Int. Conf. Mach. and Learn.}, Long Beach, California,
  USA, Jun. 2019, pp. 1182--1192.

\bibitem{song2020infomax}
Yuxuan Song, Minkai Xu, Lantao Yu, Hao Zhou, Shuo Shao, and Yong Yu,
\newblock ``Infomax neural joint source-channel coding via adversarial bit
  flip,''
\newblock in {\em Proc. AAAI Conf. Artificial Intell.}, New York, USA, Feb.
  2020, pp. 5834--5841.

\bibitem{Marchioro2020Adversarial}
Thomas Marchioro, Nicola Laurenti, and Deniz Gündüz,
\newblock ``Adversarial networks for secure wireless communications,''
\newblock in {\em Proc. Int. Conf. Acoustics Speech Signal Process.}, Virtual,
  May 2020, pp. 8748--8752.

\bibitem{Erdemir2022Privacy}
Ecenaz Erdemir, Pier~Luigi Dragotti, and Deniz Gündüz,
\newblock ``Privacy-aware communication over a wiretap channel with generative
  networks,''
\newblock in {\em Proc. Int. Conf. Acoustics Speech Signal Process.}, Shenzhen,
  China, Oct. 2022, pp. 2989--2993.

\bibitem{xu2021wireless}
Jialong Xu, Bo~Ai, Wei Chen, Ang Yang, Peng Sun, and Miguel Rodrigues,
\newblock ``Wireless image transmission using deep source channel coding with
  attention modules,''
\newblock {\em {IEEE} Trans. Circuits Syst. Video Technol.}, Apr. 2021.

\bibitem{tishby2000information}
Naftali Tishby, Fernando~C Pereira, and William Bialek,
\newblock ``The information bottleneck method,'' Available:
  \url{https://arxiv.org/abs/physics/0004057}, 2000.

\bibitem{pan2021disentangled}
Ziqi Pan, Li~Niu, Jianfu Zhang, and Liqing Zhang,
\newblock ``Disentangled information bottleneck,''
\newblock in {\em Proc. AAAI Conf. Artificial Intell.}, Virtual, Feb. 2021, pp.
  9285--9293.

\bibitem{alemi2016deep}
Alexander~A Alemi, Ian Fischer, Joshua~V Dillon, and Kevin Murphy,
\newblock ``Deep variational information bottleneck,'' Available:
  \url{https://arxiv.org/abs/1612.00410}, 2016.

\bibitem{pmlr-v80-kim18b}
Hyunjik Kim and Andriy Mnih,
\newblock ``Disentangling by factorising,''
\newblock in {\em Proc. Int. Conf. Mach. and Learn.}, Stockholm, Sweden, Jul.
  2018, pp. 2649--2658.

\bibitem{NEURIPS2018_1ee3dfcd}
Ricky T.~Q. Chen, Xuechen Li, Roger~B Grosse, and David~K Duvenaud,
\newblock ``Isolating sources of disentanglement in variational autoencoders,''
\newblock in {\em Proc. Adv. Neural Inform. Process. Syst.}, Montreal, Canada,
  Dec. 2018, p. 2610–2620.

\bibitem{Chen_2021_ICCV}
Zhi Chen, Yadan Luo, Ruihong Qiu, Sen Wang, Zi~Huang, Jingjing Li, and Zheng
  Zhang,
\newblock ``Semantics disentangling for generalized zero-shot learning,''
\newblock in {\em Proc. Int. Conf. Comput. Vis.}, Virtual, Oct. 2021, pp.
  8712--8720.

\bibitem{nguyen2010estimating}
XuanLong Nguyen, Martin~J. Wainwright, and Michael~I. Jordan,
\newblock ``Estimating divergence functionals and the likelihood ratio by
  convex risk minimization,''
\newblock {\em {IEEE} Trans. Inf. Theory}, vol. 56, no. 11, pp. 5847--5861,
  Nov. 2010.

\bibitem{Hadad2018two}
Naama Hadad, Lior Wolf, and Moni Shahar,
\newblock ``A two-step disentanglement method,''
\newblock in {\em Proc. {IEEE} Conf. Comput. Vis. Pattern Recog.}, Salt Lake
  City, USA, Jun. 2018, pp. 772--780.

\bibitem{lecun1998mnist}
Emre Sariyildiz, Haoyong Yu, and Kouhei Ohnishi,
\newblock ``Gradient-based learning applied to document recognition,''
\newblock {\em Proceedings of the IEEE}, vol. 86, no. 11, pp. 2278--2324, Nov.
  1998.

\bibitem{Zhang2017Age}
Zhifei Zhang, Yang Song, and Hairong Qi,
\newblock ``Age progression/regression by conditional adversarial
  autoencoder,''
\newblock in {\em Proc. {IEEE} Conf. Comput. Vis. Pattern Recog.}, Hawaii, USA,
  Jul. 2017, pp. 5810--5818.

\bibitem{van2008visualizing}
Laurens Van~der Maaten and Geoffrey Hinton,
\newblock ``Visualizing data using t-{SNE}.,''
\newblock {\em J. Mach. Learn. Research}, vol. 9, no. 11, Nov. 2008.

\end{thebibliography}

\end{document}